\documentclass[10pt,twoside]{article}
\usepackage{epsf}
\usepackage{graphicx}
\usepackage{lscape}
\makeindex 
\begin{document}

\title{Variable Sine Algorithmic Analysis (VSAA): A new method of frequency analysis - Applications.}
\author{S. Tsantilas\thanks{E-mail:
stsant@phys.uoa.gr} \, and H. Rovithis-Livaniou\thanks{E-mail: elivan@phys.uoa.gr}\\
Section of Astrophysics, Astronomy \& Mechanics, Dept. of Physics, \\
Athens University, Panepistimiopolis, Zografos 157 84, Athens, Greece.}

\maketitle

\begin{abstract} 
A new method of frequency analysis is presented in detail. This new method - Variable Sine Algorithmic Analysis, (VSAA) - is based on a single variable sine function and it is powered by the simplex algorithm. It is used in cases of phenomena triggered by a single mechanism, where Fourier Transform and Wavelet Analysis fail to describe practically and efficiently. Applications are given on the orbital period variation of two RS CVn type binaries: RS CVn itself and CG Cyg. With the use of the Applegate's mechanism, the variation of their subsurface magnetic field has been acquired.
\end{abstract}
\textbf{keywords:} method: data analysis--binaries: close--stars: magnetic activity

\section{Introduction} 
For decades, the only way of time series frequency analysis was the well-known Fourier transform (FT) in various forms, mainly DFT (Descrete Fourier Transform) and FFT (Fast Fourier Transform). More recently, the Wavelet Analysis (WA), appeared as a more promising method, to overcome some of the Fourier weaknesses (Daubechies, 1990). Nevertheless, a very significant drawback for both Fourier and WA, is the difficulty of assigning a unique physical mechanism to the observed time series. As an example one can take a variable magnetic field that is the only mechanism that produces the data set. In order to reconstruct the observed signal, the two previous methods require the superposition of a huge number of frequencies, some of which might have no physical meaning at all, given that the mechanism is only one. In such a case, it is very difficult to distinguish the important ones. In order to handle this type of signals, we introduce the VSAA (Variable Sine Algorithmic Analysis) method.
The main characteristics of the new method are:\\
1.	Instead of decomposing the time series to a large number of frequencies it actually decodes \textit{algorithmically} the signal and produces only one variable frequency. It is very important here to make clear that this is a completely different approach compared with FT and WA: There are no series of orthogonal functions with constant periods, nor convolution with wavelet functions, but instead only one variable sine function that describes the variability of the signal through an iterative procedure. In the case of a single mechanism, VSAA is the only method that gives physically compatible results.\\
2.	In the classic Fourier transform, the analysis window is the complete range of the data. So, the results are greatly affected by this range. VSAA uses the 'sliding window' technique just like in the wavelet analysis and the short-time FT, for extracting local-frequency information from the signal. Therefore, the frequency variation is time dependent and describes much more efficiently the observed data.\\
3.	It includes the feature of filtering. The range of the sliding window determines the scaling of the analysis: A small window inserts a high band pass filter that can trace high frequencies. A large window inserts a low band pass filter that can trace low frequencies.\\
4.	It encloses adjustable accuracy threshold. \\
5.	It can handle without any modification non-equispaced data. Although there are some recent attempts to overcome this problem (Palmer \& Smylie 2004, Foster 1996, Cai \& Brown 1998), in most cases equispaced data is a prerequisite for both classic FT and WA.\\
6. It is powered by an arithmetic method, i.e. simplex (Kallrath, 1987). Every improvement on this method can be immediately included in the code and make the procedure more efficient. 

\section{The method}
VSAA is based on the idea of a single variable frequency (Tsantilas \& Rovithis, 2005). In many physical phenomena, it is known (or suspected) that the cause behind the time series signal is a single mechanism. In such a case, the classical approaches (FT and WA) fail to explain the mechanism because of the large number of frequencies required to reproduce the signal. On the other hand, VSAA decodes the signal, producing a single frequency that varies in time, reflecting the variations in magnitude of the driving mechanism.
VSAA uses the function
\begin{equation}
f(t)=a \cdot sin(bt+c),
\end{equation}
where $a=a(t), b=b(t)$ and $c=c(t)$, i.e. they are functions of time. Then, the given data set will be fragmented and each part will be fitted by this function using the simplex algorithm. The procedure is described in detail in the following subsections.

\subsection{Preparation of the data set.}
Lets assume a time series Q in the form of
\begin{equation}
Q: (t_i,q_i), i=1...N,
\end{equation}
where $N$ is the number of the data.
There is no need for the data to meet any special requirements, i.e. to be equispaced. The only thing is that given the form of the function (1), the signal should be more or less symmetrical to the zero value. This can be easily obtained by fitting it by a linear function
\begin{equation}
y=kt+m,
\end{equation}
which represents the long term trend of the time series. Then, by subtracting\\
$$s_i=q_i-y=q_i-kt_i-m,$$
we acquire the proper signal
$$S:(t_i,s_i), i=1...N.$$

\subsection{Inputs and filtering}
The time series with the above modification (if needed) will be used as input for the code. VSAA provides the ability to the user to define the filter of the signal analysis by adjusting the width $n$ of the 'sliding window'. The width of the 'sliding window' is the number of the data that will be fitted in each step $j$.
Next, the starting values of $a,b,c$ parameters should be given, along with the accuracy threshold. Also, the number of iterations can be adjusted in this stage.

\subsection{Fitting of the signal}
In every step $j$, the simplex algorithm fits a 'window' W of data
$$W:(s_j,s_{j+n-1}), j=1...N-n,$$
the width $n$ of which has been defined by the user in the filtering stage of the inputs. When the algorithm reaches the desired accuracy or the number of iterations, it produces the vector
\begin{equation}
v_j=(t_j,a_j,b_j,c_j,\sigma_j),
\end{equation}
where $a_j=a_j(t_j),b_j=b_j(t_j), c_j=c_j(t_j)$ are the coefficients of the variable sine function and $\sigma_j=\sigma_j(t_j)$ is the error of the $j$ fit.
Then, the code moves to the next step $j+1$, using the window
$$(s_{j+1},s_{j+n}),$$
 with the same width n, and repeats in the same way.  The previously acquired $a_j,b_j,c_j$ are used as starting values for the fitting procedure of the variable sine function (1). Notice that \textbf{the next step is only one point further from the previous}. From the above, a more or less continuous form for the $a_j=a_j(t_j),b_j=b_j(t_j),$ and $c_j=c_j(t_j)$  can be achieved.

\subsection{Outputs}
When the sliding window has run through the whole data set, the code produces a set of vectors
$$v_j=(t_j, a_j,b_j,c_j,\sigma_j), j=1...N-n,$$
where\\
$a_j(t)=\Delta f(t):$ the amplitude of the variable frequency $f(t)$,\\
$\frac{b_j(t)}{2\pi}=f(t):$ the variable frequency,\\
$c_j(t):$ the phase shift of the function (1),\\
$\sigma_j=\sqrt{\frac{\Sigma(s_j-a_j)^2}{n}}$.\\
The signal can be then reproduced by the variable sine function
$$s_j \approx a_jsin(b_jt_j+c_j), j=1...N-n.$$

\section{Applications to binary stars' period changes}
Some of the close binary systems are known to exhibit solar-type activity to a much stronger degree than our Sun. Such are the contact binaries of \textit{W UMa}-type, the \textit{LMXBs}, the \textit{CVs}, the \textit{Algols} and the \textit{RS CVn}-type systems.  Their activity seems to be the result of their rapid rotation and the deep convective zone of one (or both) of their components.  
The observed orbital period changes of these systems, were connected to their magnetic activity long ago, (Matese \& Whitmire 1983); an idea that was widely accepted.  And now many periodic or quasi-periodic changes of the orbital period of close binary systems are considered as being the result of magnetic activity cycles similar to those of our Sun, because the structure of a magnetically active star varies during an activity cycle, (Applegate \& Patterson 1987).  These variations change the quadrupole moment of the active star. And as the star expands and contracts, the equipotential surfaces -in the Roche geometry used- are deformed.
Moreover, according to Applegate (1992) the produced period modulations can be explained by the gravitational coupling of the orbit to deform the shape of the magnetic component.  And this deformation exerts a periodic torque on the orbit that modulates the orbital period of the system.
Assuming that the observed orbital period changes of a close binary are due \textbf{\textit{only}} to the magnetic activity in one of its components we propose a \textbf{\textit{new approach}} using the \textbf{VSAA} method, that enables us to get some further information about this star's magnetic field; especially its variation during a magnetic cycle.  Furthermore, it enables us to explain the observed quasi-periodic cycles, as well as the great differences in the 'periodic' cycle values derived by the various investigators.
According to Applegate (1992), a constant magnetic field acting in one of the components of a binary, can produce a periodic modulation of the period function \textbf{\textit{P(E)}} of amplitude \textbf{\textit{$ \Delta P$}} and periodicity \textbf{\textit{$P_{mod}$}}. The magnetic field's intensity is then estimated by the formula:
\begin{equation}
B^{2} \approx 10\frac{GM^2}{R^4} \left (\frac{a}{R} \right)^2 \frac{\Delta P}{P_{mod}},
\end{equation}
where all symbols have their usual meaning; so: $M$ and $R$ denote the mass and the radius of the active star respectively, $a$ stands for the two components separation, and $B$ is the intensity of the magnetic field.
From the foregoing formula, it is obvious that constant $ \Delta P$ and $P_{mod}$ values yield to a constant $B$. But the findings of the (O-C) diagrams analyses show that the P(E) functions \textbf{do not exhibit a strictly periodic form, and the $ \Delta P$ and $P_{mod}$ are time depended}, (e.g. Rovithis-Livaniou 2005). Hence, \textbf{the magnetic field is variable}; and its variability could be found if it was possible to estimate $P_{mod}$ and $ \Delta P$ variability.
So, starting with the (O-C) diagram analysis of a binary -using Kalimeris et al. (1994) method- it is very easy to get its corresponding period variations. They are usually expressed in terms of the 
$$P_j=P_j(E_j)-P_e,$$
function, where $P_e$ is the constant ephemeris period used for the (O-C) diagram construction of the system under examination. The acquired time series
$$S:(E_j, P_j),$$
is used as input for the VSAA method. Notice that the reader must not confuse the period $P_j$, which serves as signal data in the time series, with the $P_{mod,j}= \frac{1}{f_j(E_j)}$ acquired from VSAA, which denotes the period of the variability of $P_j$.
Working in this way we manage to have \textbf{a series of values} for the $P_{mod}$ and $ \Delta P$ instead of having constant values for the whole time interval for which observational data exist. That is, to have both of them as function of the epoch $E$; i.e. to have them as functions of time. Having achieved this, assuming that the observed period changes of the close binary under examination are due to magnetic activity only, and keeping Applegate's formalism, it is easy to deduce the magnetic field's intensity as a function of time and estimate how it varies during the interval for which observational data exist.\\
The foregoing described new approach was applied to two very active binaries of RS CVn-type; namely to the RS CVn itself, and to CG Cyg.  These two systems were purposely chosen, as: 1) both have been observed for more than a hundred years, and 2) CG Cyg belongs to the short-period sub-class of RS CVn-type binaries, $P \approx 0.63$ days; while the period of the first is more typical for its type, $P \approx 4.8$ days.
Although a big number of light curve analyses and corresponding data exist for both systems, we are not paying special attention, as they are out of the present study's interest. So, giving only some very general information about the systems, we continue with the construction of their (O-C) diagrams and their subsequent analysis in order to find their period changes. This is carried out using one of the modern techniques, i.e. the first continuous method, (Kalimeris et al. 1994).  Finally, we apply the proposed new approach to their period changes as they are expressed via their corresponding $P(E)-Pe$ functions.

\begin{figure}[ht!]
\centerline{\includegraphics[width=8cm]{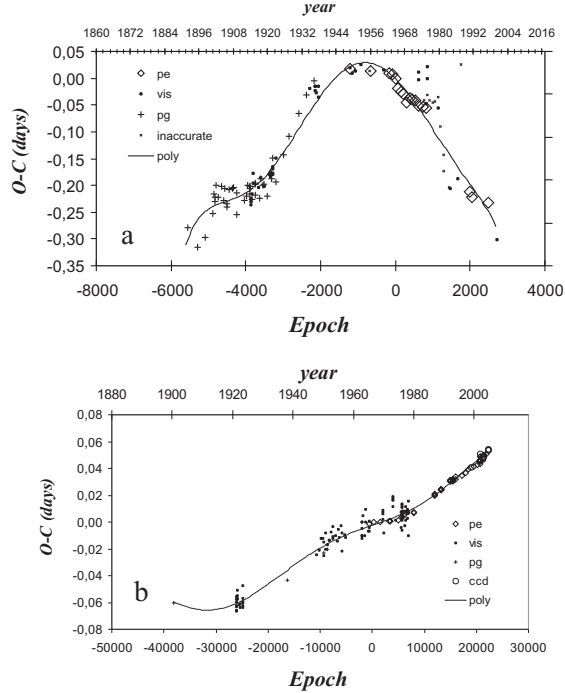}}
\caption[]{a. The (O-C) diagram of RS CVn, where the continuous line denotes its best description.\\
b. Similar as Fig.1a but for CG Cyg.}  
\end{figure}

\begin{figure}[ht!]
\centerline{\includegraphics[width=8cm]{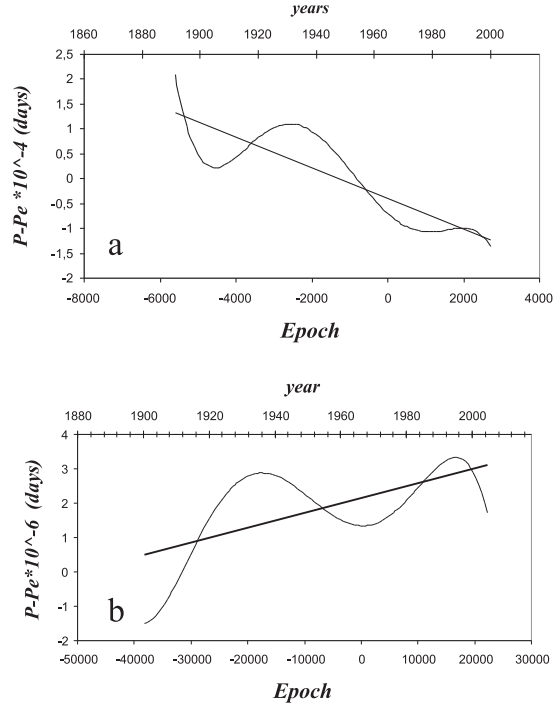}}
\caption[]{a. The P(E)-Pe function of RS CVn, where the heavy straight line denotes the detected long-term period variation.\\
b. Similar as Fig. 2a, but for CG Cyg.}  
\end{figure}

\subsection{About the Candidates}
The RS CVn Binary:\\
The variable star RS CVn was discovered and identified as an Algol-type binary by Ceraski (1914). Its light elements, light curve nature, and components' classification attracted the interest of many investigators, and Hall (1976) proposed it as the prototype of a new class of close binaries. The basic characteristics of this new category are: high chromospheric activity, light curve asymmetry, presence of stellar spots, and migrating photometric waves.
RS CVn (BD +36$^{\circ}$2344) is a short period, P=4.7978d, eclipsing binary consisting of a G9 IV and an F4 components, with effective temperatures of 5090 and 6560 $^{\circ}$K, respectively (Popper, 1988). From the large number of light curves and analyses for RS CVn, in our subsequent analysis the absolute elements given by Eaton et al. (1993) will be used.\\
The CG Cyg Binary:\\
CG Cyg (BD +34$^{\circ}$4217) is a short period, P=0.6311d, eclipsing binary consisting of a G9.5V and a K3V components, (Naftilan \& Milone, 1979), with effective temperatures of 5200 and 4400 $^{\circ}$K, respectively (Kjurkchieva et al. 2003). It has a variable orbital period (Milone \& Ziebarth, 1974) and also variable mean brightness outside eclipses,  (Sowell et al. 1987). Hall (1991) argued that the chromospheric activity is due to a subsurface magnetic field triggered by the Applegate mechanism. This explanation seems to agree also with its brightness variation. And Lazaro \& Arevalo (1997) discovered emission excess in the $H_a$ line for both of its stars.  From the large number of light curves and analyses for CG Cyg, in our subsequent analysis the absolute elements given by Kozhevnikova et al. (2005) will be used.

\subsection{Orbital Period Analysis of the Candidate Binaries}
Taking into account all times of minimum light for the two systems, their (O-C) diagrams were constructed, and are presented in Figs. 1(a\&b), respectively. More explicitly: Fig. 1a is for RS CVn and is based on the light elements (Rodono et al., 1995):
$$Min I = 2438889.3300 + 4.797817 E.$$
Fig. 1b, presents the (O-C) diagram of CG Cyg, and was made with the aid of (Milone \& Ziebarth, 1974):
$$Min I = 2439425.1221+0.631141 E.$$  
Then, the (O-C) diagrams of the systems were analysed using Kalimeris et al. method. According to this -and giving different weights to the various kinds of the individual observational points- the appropriate higher order polynomials for their best description were found.   The continuous lines in Figs 1(a\&b) denote them, while their coefficients are given in Table 1.

\begin{table}[ht!]
\caption{Coefficients of the polynomials used for the (O-C) diagrams description of the systems}
\smallskip
\begin{center}
{\scriptsize\begin{tabular}{crr}
\hline
\noalign{\smallskip}
Coefficient & RS CVn & CG Cyg\\
\noalign{\smallskip}
\hline
\noalign{\smallskip}
$c_{0}$ &  $-0.00097$ & $-0.002376775$\\
$c_{1}$ &  $-0.4185$ & $ 0.053702500$\\
$c_{2}$ &  $-1.23894$ & $-0.005583185$\\
$c_{3}$ &  $2.008667$ & $ 0.263103001$\\
$c_{4}$ &  $2.324922$ & $ 0.126889974$\\
$c_{5}$ &  $-4.28911$ & $-0.443503916$\\
$c_{6}$ &  $-4.22124$ & $-0.302704433$\\
\noalign{\smallskip}
\hline
\noalign{\smallskip}
\multicolumn{3}{c}{Scale constant of RS CVn: $6 \times 10^3$}\\
\multicolumn{3}{c}{Scale constant of CG Cyg: $4 \times 10^4$}
\end{tabular}
}
\end{center}
\end{table}

\noindent Afterwards, the period variations of the two RS CVn-type binaries were calculated and are presented in Figs. 2(a\&b), respectively. The orbital period changes are in the form of P(E)-Pe functions, where Pe is the constant period used for the (O-C) diagram construction of each one of the systems. And the straight lines in Figs. 2(a\&b) denote the detected \textbf{long-term variations} of the period of the systems.
For the RS CVn this was found to be equal to: 
$-3.0866 \times 10^{-8} d/E$ that corresponds to $-0.2028 s/y$.
For the CG Cyg it was found to be equal to: 
$4.3362 \times 10^{-11} d/E$ corresponding to $0.0021 s/y$. 
So, we notice that RS CVn shows very large long-term variation; actually it is 2 orders of magnitude greater than that of CG Cyg.\\
\noindent After subtracting the detected secular period decrease for RS CVn and increase for CG Cyg, their P(E)-Pe functions show a wave like variation. On this data set, a search for periodicities was made, using FT and WA. For RS CVn, from FT we got three major periodicities (among a number of others with smaller amplitudes), $P_{mod,1}$=109y, $P_{mod,2}$=54.5y and $P_{mod,3}$=36.3y with corresponding amplitudes of the same order of magnitude, $\Delta P_1$=4.20 $\times 10^{-5}$=3.62880s, $\Delta P_2$=4.32$\times 10^{-5}$d=3.73248s, $\Delta P_3$=1.60$\times 10^{-5}$d=1.38240s. We superimposed these periodicities in order to reproduce the data. Apart from the inability of physically interpreting the results, one can notice the weak fit on the P(E)-Pe function (Fig. 3a). The same holds for WA analysis (Torrence \& Combo, 1998), which can be seen in Fig. 4. For CG Cyg, Fourier transform analysis resulted to three major periodicities: $P_{mod,1}$ of 104.72y with amplitude equal to $\Delta P_1=8.00 \times 10^{-7} d=0.06912 s$; $P_{mod,2}=52.36 y$ with amplitude equal to $\Delta P_2=9.30 \times 10^{-7} d=0.08035 s$; and $P_{mod,3}=34.91 y$ with amplitude $\Delta P_3=3.85\times 10^{-7} d=0.03326 s$. The Fourier reconstructed P(E)-Pe function is presented in Fig. 3b. The Wavelet Analysis can be seen in Fig. 5.

\begin{figure}[ht!]
\centerline{\includegraphics[width=8cm]{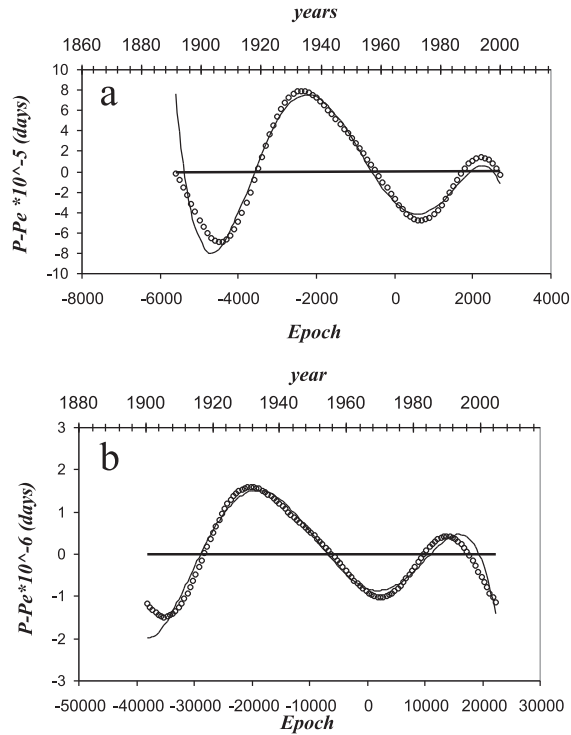}}
\caption[]{a.  The P(E)-Pe function of RS CVn (continuous line) and the combined fit (circles) of the three periodic terms acquired with the Fourier analysis.\\
b.  Same as Fig. 3a, but for CG Cyg.}  
\end{figure}

\begin{figure}[ht!]
\centerline{\includegraphics[width=12cm]{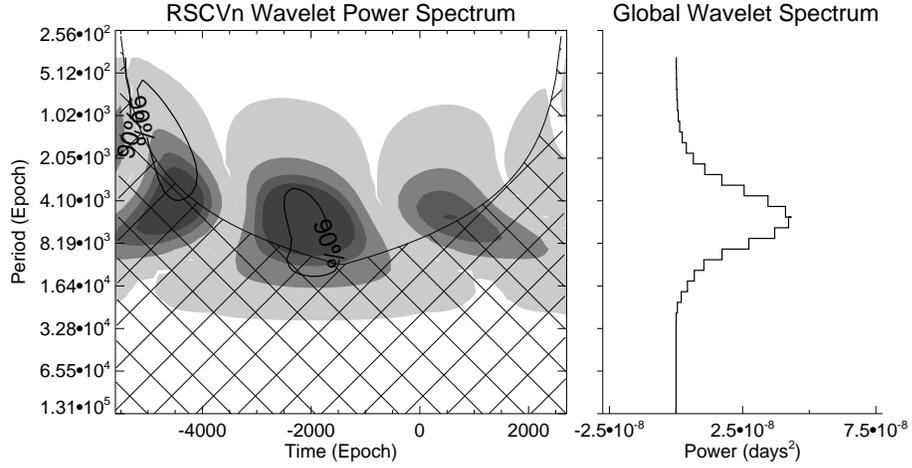}}
\caption[]{Wavelet analysis of RS CVn, using 'Mexican hat' (DOG, m=2) wavelet. Before $-4000 E$ there is a major feature centered at about $4.1 \times 10^3 E \approx 54$ years with a power of $4.5 \times 10^{-8} days^2$. After $-4000 E$, there are two major features centered at about $6.5 \times 10^3 E \approx 85$ years with a slight lower power. Grayscale filled contours corresponds to powers: $10^{-10}, 10^{-8}, 3 \times 10^{-8}$ and $5 \times 10^{-8} days^2.$ Wavelet software for Figs 4 and 5 was provided by C. Torrence and G. Compo and is available at URL: http://paos.colorado.edu/research/wavelets/.}  
\end{figure}

\begin{figure}[ht!]
\centerline{\includegraphics[width=12cm]{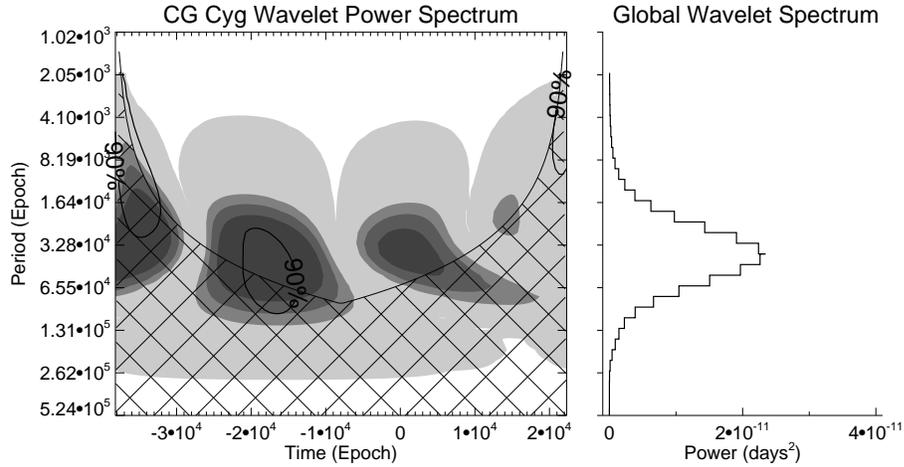}}
\caption[]{Wavelet analysis of CG Cyg, using 'Mexican hat' (DOG, m=2) wavelet. Three main features are centered at about 3, 5 and 3.5 $\times 10^4 \,E$ corresponding to 52, 86 and 60 years with a power of $2 \times 10^{-11} days^2$. Grayscale filled contours corresponds to powers: $10^{-14}, 5 \times 10^{-12}, 10^{-11}$ and $2 \times 10^{-11} days^2$.}  
\end{figure}

\begin{figure}[ht!]
\centerline{\includegraphics[width=8cm]{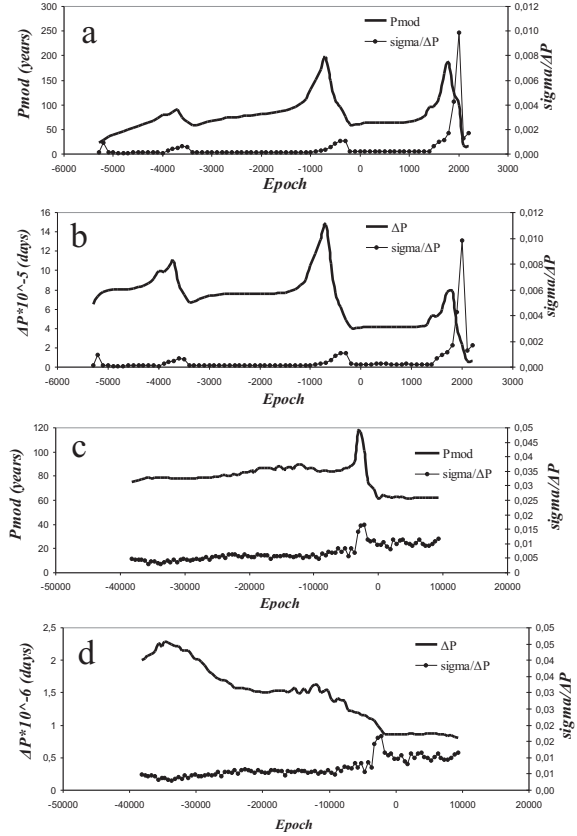}}
\caption[]{a-b. The $P_{mod}$ and $\Delta P$ variation of RS CVn.\\
c-d. The $P_{mod}$ and $\Delta P$ variation of  CG Cyg.}  
\end{figure}

\begin{figure}[ht!]
\centerline{\includegraphics[width=8cm]{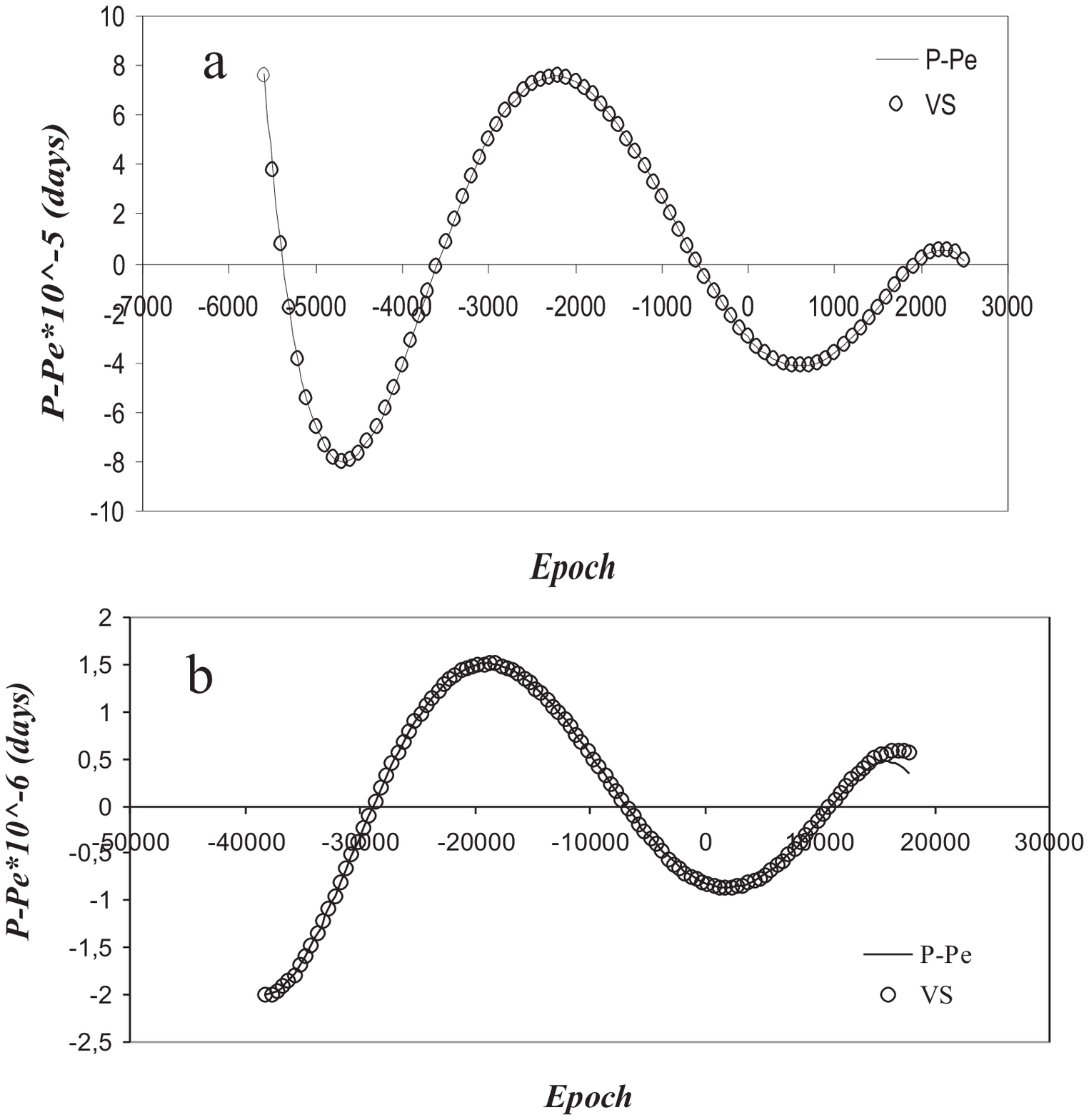}}
\caption[]{a. The wave like changes of the orbital period changes of RS CVn, after subtracting the long-term variation. Circles denote the Variable Sine fitting.\\
b.  Same as Fig. 7a, but for CG Cyg.}  
\end{figure}

\noindent On the other hand, using the same data set as input, we applied the \textbf{VSAA} code. From this procedure, a set of $P_{mod,j}=1/f_j(E_j)$ and $\Delta P_j(E_j)=a_j(t)=\Delta f(t)$ (as described in subsection 2.4 and section 3) have been acquired (Figs. 6a-d). Taking each system separately we found:\\
a) RS CVn: The VSAA analysis showed a variable $P_{mod}$ (Fig. 6a), which takes values around 60-70 years in good agreement with most of the published periodicities. There are three regions ((-4000,-3400), (-1100,-200) and (1400-2200)) corresponding to much larger values. This could be partly an artefact, imputed to small discontinuities as the large sigma on these regions implies.  $\Delta P$ (Fig. 6b) describes quite accurately the amplitude of the variable periodic term. Reconstructing the P(E)-Pe function from the acquired $P_{mod}$ and $\Delta P$, we get the excellent fit presented in fig. 7a.\\
b) CG Cyg:  From the VSAA we can see that $P_{mod}$ for the (-38200, -5200) time interval takes values around 80 years. After a fast drop, it presents a variable periodicity around 65 years for the last time interval (Fig. 6c). In addition, Fig. 6d presents the amplitude of the variable periodic term. From the reconstruction of the P(E)-Pe function from the acquired $P_{mod}$ and $\Delta P$, we have the fit presented in fig. 7b.\\
In order to estimate the variation of the binary's magnetic field, VSAA provide us this unique opportunity. Using equation (5), \textbf{we can compute the subsurface magnetic field of the active component of each star}. Fig. 8a presents in detail the variation of the magnetic field's intensity of the active component of the RS CVn binary, assuming that \textbf{its observed period variations are due to magnetic activity only}, and there is not any other mechanism that could produce with its action a change in the orbital period of this system. In Fig. 8b this variation is shown for the primary component of CG Cyg.
\begin{figure}[ht!]
\centerline{\includegraphics[width=8cm]{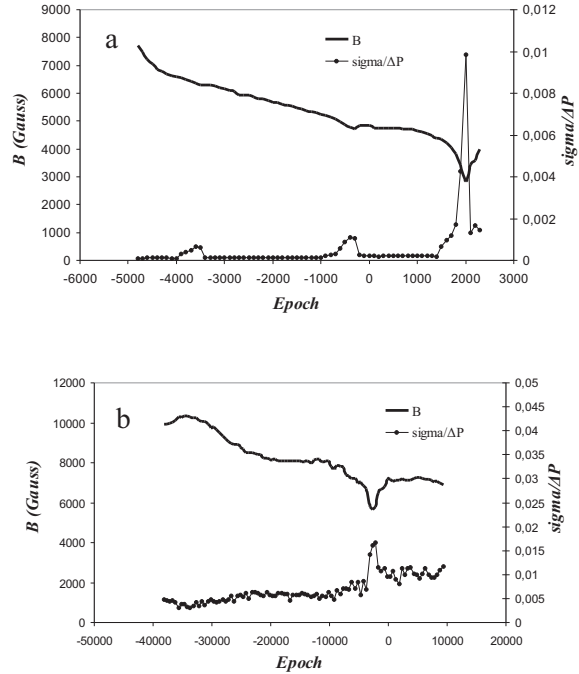}}
\caption[]{a. The subsurface magnetic field variation of the primary component of RS CVn.\\
b.  Same as Fig. 8a, but for CG Cyg.}
\end{figure}

\section{Discussion}
The purpose of the VSAA method \textbf{is not to substitute FT and WA}, but to be used in situations of single mechanisms where the previous methods give physically questionable results. So, there is no question of comparison, because in these cases VSAA is an obligatory choice, while in the other cases FT and WA are much more efficient and reliable. Nevertheless, we will try to present some advantages and disadvantages of VSAA with respect to FT and WA because these are the only points of reference we have.\\
Advantages:\\
1.	Obviously, the more rational handling of single-mechanism signals.\\
2. The simplicity in the use of the code and the straight-forward interpretation of the results.\\
3.	In the classic FT, frequencies are not sensitive to time. In WA they are, but the almost chaotic amount of information makes the interpretation of the results extremely difficult. In VSAA on the other hand, frequencies are sensitive to time and in addition it provides a clear description of the signal, transferring the point of interest to the time variance of the frequency $f$ and its amplitude $\Delta f$.\\
4.	It can handle without any modification non-equispaced data (although the algorithm can work more efficiently with equispaced data). In most cases this is a prerequisite for both classic FT and WA.\\
5.	The method is \textit{algorithmic}. The core of the code is an arithmetic method, i.e. simplex, so every improvement on this method can be immediately and easily included in the code and make the procedure more efficient.\\
Disadvantages:\\
1.	The need of well-defined starting values.\\
2.	The dependence of an adequate accuracy threshold.\\

\noindent Concerning the applications: Applegate analysed two mechanisms by which magnetic activity can change the quadrupole moment of a star and he found that very small changes in the luminosity, $dL/L \sim 0.1$, and in the star's differential rotation $d \Omega / \Omega \sim 0.01$, can explain orbital period changes of the order of $dP/P \sim 10^{-5}$, which are typical for close binaries. Furthermore, Applegate's mechanism has several testable predictions, which according to some investigators are fulfilled, while according to some others they don't (Lanza \& Rodon\`{o}, 2002).
More recently, R\"{u}diger et al. (2002) estimated the magnetic field intensity needed to produce the required quadrupole moment changes of RS CVn-type binaries.
In the two examined cases of RS CVn and CG Cyg, \textbf{there are not indications of the presence of a third component}.  Therefore, \textbf{such a possibility was excluded}, as causing their detected orbital period changes. Moreover, RS CVn-type systems are detached binaries and \textbf{mass transfer} between the two components \textbf{should be ruled out} too, as another possible explanation of their observed period variations.  So, our assumption that the observed orbital period variations of these two binaries are the result of their magnetic activity only, should be correct.\\
From the (O-C) diagram analysis of the binaries, performed by Kalimeris et al method, we acquired the P(E)-Pe functions. This is necessary in our subsequent analysis, and cannot be achieved with the traditional analysis methods (Rovithis-Livaniou, 2001). The P(E)-Pe functions were used as input for the VSAA algorithm in order to get the $P_{mod}$ and $\Delta P$ time variation. Then, from Applegate's theory, we were able to estimate the subsurface magnetic field of the star's components.\\
The magnetic field of 2.8-7.7 kG calculated for the RS CVn primary component requires a $dL/L$ of  0.04-0.92. Although the lower boundary is in good agreement with the Applegate's model, for the upper boundary we have a clearly strained fit. On the other hand, only a $dL/L \sim 0.02-0.38$ is needed for the 5.6-10.3 kG CG Cyg primary's star magnetic field (figure 8b), which is in good agreement with the model. As there are indications of magnetic activity on the secondary component too, (Lazaro \& Arevalo, 1997), we applied also the same procedure to this component. The resulting subsurface magnetic field varies between 9.2 and 16.9 kG. This variation requires a $dL/L \sim 0.43-3.27$, which is 4 to 32 times larger than that predicted by the model. Thus, one can conclude that the period modulation of CG Cyg due to magnetic mechanism, should be probably assigned to magnetic activity of the primary component mostly.\\

\noindent \textit{The VSAA code, manual and examples, can be downloaded via http at\\
http://users.uoa.gr/$\sim$stsant.}\\

\noindent \textbf{Acknowledgements.} This work was financial supported by the Athens University (grant 70/4/3305).

{}

\end{document}